\documentclass[twocolumn,amssymb,showpacs,aps,prl]{revtex4-1}
\usepackage{}
\usepackage{stmaryrd}
\usepackage{amsmath}
\usepackage{amssymb}
\usepackage{float}
\usepackage{textcomp}
\usepackage{graphicx}
\usepackage{titletoc}
\usepackage{booktabs}
\usepackage{array, color}
\usepackage{tabularx}
\usepackage{latexsym,bm}
\usepackage{graphics}

\begin{document}

\title{High-harmonic and terahertz wave spectroscopy (HATS) for aligned molecules}

\author{Yindong Huang$^{1}$, Chao Meng$^{1}$, Jing Zhao$^{1}$, Xiaowei Wang$^{1}$, Zhihui L\"u$^{1}$, \\
Dongwen Zhang$^{1}$, Jianmin Yuan$^{1,2}$}
\email{jmyuan@nudt.edu.cn}
\author{Zengxiu Zhao$^{1}$}
\email{zhao.zengxiu@gmail.com}
\address{$^1$Department of Physics, College of Science, National University of Defense Technology, Changsha 410073, Hunan, People's Republic of China\\
$^2$IFSA Collaborative Innovation Center, Shanghai Jiao Tong University, Shanghai 200240, People's Republic of China}

\date{\today}

\begin{abstract}
We present the experimental and theoretical details of our recent published letter [Phys. Rev. Lett. 115. 123002] on synchronized high-harmonic and terahertz-wave spectroscopy (HATS) from nonadiabatically aligned nitrogen molecules in dual-color laser fields.
Associating the alignment-angle dependent terahertz wave generation with the synchronizing high-harmonic signal, the angular differential photoionization cross section (PICS) for molecules can be reconstructed, and the minima of the angle on PICS show great convergence between the theoretical predictions and the experimental deduced results.
We also show the optimal relative phase between the dual-color laser fields for terahertz wave generation dose not change with the alignment angle at a precision of $50$ attoseconds.
This all-optical method provides an alternative for investigating molecular structures and dynamics.

\end{abstract}


\maketitle

\section{I. Introduction}
Recent advances in laser science have shed new light on the real-time observation and manipulation on atomic-scale motion of electrons~\cite{Kraus2009}.
Taking advantage of the quantum coherence of ultrafast dynamics, ionization, vibrational excitation, band transition and resonance, have been traced via modulating the interference between the related quantum pathways for atoms~\cite{Dudovich2006, ZhangDW2012, Pedatzur2015,ZhaoJ2015}, molecules~\cite{Kanai2005, Meckel2014} or solids~\cite{Vampa2015,Hohenleutner2015}.

Complete measurements are desired to obtain the full aspects of dynamics in strong laser fields. So far more than one physical quantities have been measured in coincidence such as ionic fragmentation, electrons or photons. In our previous work, we have demonstrated that synchronized high-harmonics and terahertz-wave spectroscopy (HATS)~\cite{ZhangDW2012, LvZH2012,HuangYD2015}, provides the correlated information of terahertz wave generation (TWG) and high-harmonic generation (HHG), which have timescales differed by more than 6 orders of magnitudes.
In the HATS setup, a weak second harmonic field is applied to detune the HHG and TWG from gases by the fundamental pulse. The interference of the radiation between the consecutive half-cycle of the fundamental laser field is controlled by varying the delay between the two pulses. It is found that the second harmonic helps generating even-order harmonics~\cite{Dudovich2006} and dramatically enhancing the TWG~\cite{Cook00}.  By monitoring the modulation of HHG and TWG, the information on both radiation and the underlying dynamics can be identified.

Utilizing the HATS technique, the unified picture for TWG and HHG, originating from the rescattering model~\cite{Lewenstein1994}, has been drawn by taking account of the Coulomb focusing effects on the continuum electrons~\cite{ZhangDW2012,LvZH2012}. The freed electrons may directly escape or being rescattered from the parent core in the oscillating laser field.
It has been shown that the rescattering can be further classified as `hard collision' or `soft collision', where the former contributes to HHG while the latter dominates TWG.

According to the quantitative rescattering (QRS) theory~\cite{LeAT2009}, alignment-dependent HHG from molecules can be viewed as the product of the angular ionization rates and the recombination matrix elements. For the electrons contributing to TWG, the long terahertz wavelength leads to the less dependence on the spatial distributions after ionization, thus making it possible to calibrate the angular ionization probabilities from alignment-angle dependent TWG~\cite{ChenWB2015}.
Therefore, the angular differential photoionization cross section (PICS) can be deduced from the synchronizing observed angular HATS, due to the principle of the detailed balance between the angular recombination cross section and the PICS.
Recently, this idea has been performed in reconstructing the angular PICS of nitrogen molecule, which shows a great consistence between our newly experimental results~\cite{HuangYD2015} and the theoretical calculations~\cite{JinCheng2012}.

In this paper, we provide more details on HATS technique, in particular, the experimental and theoretical details on how to reconstruct the PICS of nitrogen molecules. Meanwhile, we also show that the optimal dual-color relative phase (DRP) for TWG does not change at different alignment angles in a precision of about $50$ as. The DRP induced modulations for terahertz yields from aligned, antialigned and isotropic nitrogen are all similar to the results from argon atom. Therefore, more precise time domain modulation is required to directly observe the differences of the Coulomb focusing between molecules and atoms~\cite{ChenWB2015}.

This article is organized as follows. In section II, we describe our experimental setup in details. The theory is presented in section III. In section IV, we show the experimental results and the corresponding discussions. Section V contains the conclusions and outlooks.

\section{II. Experimental Technology}
The schematic of the HATS is sketched in FIG.~1. Utilizing the dual-color fields on aligned molecules, HHG and TWG can be simultaneously measured in one run. The detailed light path has been reported~\cite{HuangYD2015}.

The experimental setup consists of a $790$~nm, $25$~fs, $1.5$~mJ, $1$~kHz Ti:sapphire laser (Femtolasers) and vacuum chambers for HATS.
The laser beam is split into three pulses for preparing aligned molecules, generating HATS and recording terahertz waveform, respectively.

The alignment pulse is to prepare the transiently aligned molecules.  A half wave plate is inserted in this light path to change the polarization, in order to vary the alignment angle of molecules. A beam shutter is also placed to continuously turn the alignment pulse on or off, monitoring the long-term stability of the laser intensity and real-time comparing the yields with or without the alignment pulse. By dividing the signals with to the ones without the alignment pulse, the relative intensity of HHG or TWG can be obtained, improving the signal-to-noise rate for long operation of laser.
The generation pulse passes through a $30~\mu$m $\beta$-Barium Borate crystal (type I) to produce its second harmonic. The group velocity dispersion of the dual-color field is compensated by a calcite plate and their polarizations are rotated to be parallel by a dual-wavelength wave plate. Owing to the minor differences on the phase velocity between the fundamental pulse and its second harmonic, minor manipulation (about 50 as) on the DRP can be fulfilled by precisely adjusting the transmission length of a pair of fused silica wedges.
The time delay ($t_D$) between the alignment pulse and the generation pulse is controlled by a stepper motor (Newport M-MTM250PP).

The alignment pulse and generation pulse are focused $\sim$0.2~mm below and 2~mm before the orifice of the continuous gas jet (0.2~mm in diameter), whose intensities in the interaction area are estimated to be $0.7\times10^{14}$~W$\textfractionsolidus$cm$^{2}$ and $1.5\times10^{14}$~W$\textfractionsolidus$cm$^{2}$, respectively. A concave coated silver mirror is applied in focusing the dual-color generation pulse.
The jet, prepared by a laser-cut capillary, provides a supersonic expansion of gas with 1 bar backing pressure. The background pressure in the vacuum chamber is typically $10^{-3}$~mbar with gas supply ($10^{-5}$~mbar without).

A hole-drilled off-axis parabolic mirror reflects the TWG and leaks the HHG at the same time.  The harmonics pass through the hole and are recorded by a homemade spectrometer containing a flat field grating and a X-ray CCD camera. The TWG is collinearly focused with the terahertz wave detection pulse though a 1 mm thick (110)-cut ZnTe crystal for electro-optic sampling (EOS) the terahertz waveform. The typical high-harmonic spectra and terahertz waveforms from aligned and antialigned nitrogen molecules are present in FIG.~2. Each high harmonic spectrum is acquired by accumulating from 10000 laser pulses. It takes 300 ms integration time to obtain one point of the terahertz waveform and the typical dynamic range on terahertz detection in our experiment is ranging from 100 to 1000 dB. Though we do not know the absolute flux of different order of high harmonics, the same order harmonic shares the same transition probability of aluminum foil, same reflection index of grating and same quantum efficiency of CCD, which makes it possible to compare the angular or temporal properties of each single order of harmonic.

In experiment for reconstructing the PICSs of nitrogen, we keep the cutoff at harmonic order 25th and only recorded the harmonics from the 21st to 25th orders, and the relative phase between the dual-color field is settled to maximize the yield of TWG. While, to trace the optimal phase of TWG, we start from the the same relative phase of the dual-color field, then monitor the modulations of terahertz wave yields for aligned, antialigned and isotropic nitrogen, as well as the optimal phase of argon for comparison.


\begin{figure}
\includegraphics*[width=3.2in]{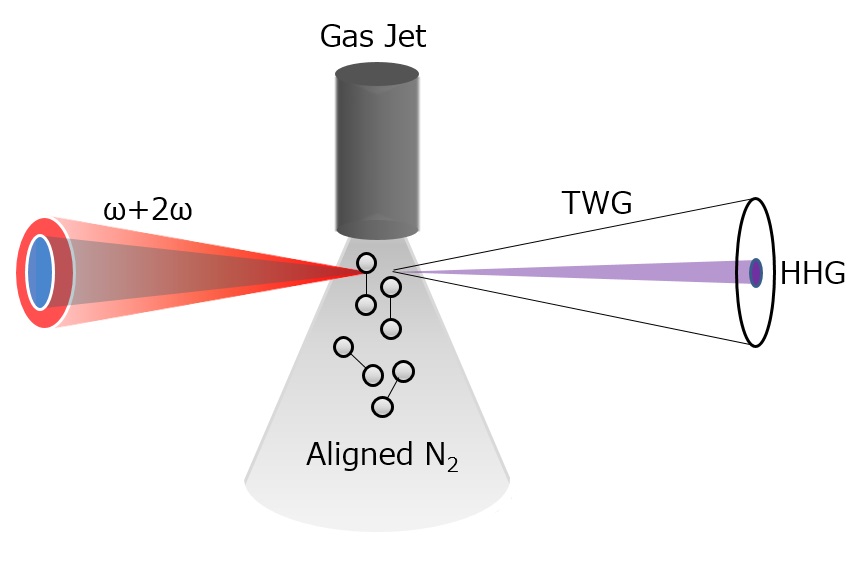}
\caption{\label{exp} Schematic of the HATS from aligned molecules. }
\end{figure}

\begin{figure*}
\includegraphics*[width=6in]{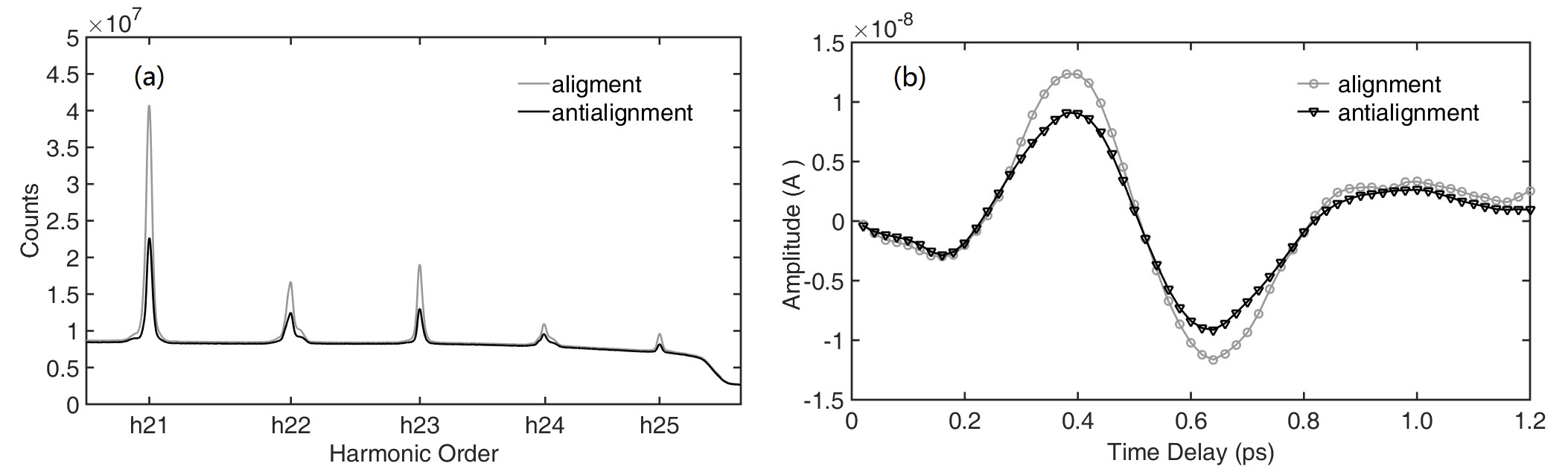}
\caption{\label{} High-harmonic spectra (a) and terahertz waveform (b) recorded at alignment (grey lines) and antialignment (black lines) moment of nitrogen. The grey cycles and the black triangles in (b) are the data points acquired from the EOS method.}
\end{figure*}

\section{III. Theory}
With the assumption of the Born-Oppenheimer approximation, the total wave packet can be divided into the independent electronic and nuclear parts. The nuclear movement include the forms of vibration and rotation, while in this paper we only consider the rotation motion.
This assumption has been widely applied in calculating the high harmonics from aligned molecules.
It is worth noting that the interaction between terahertz wave and molecules may affect the rotational wave packet for polar molecules~\cite{Fleischer2011,Egodapitiya2014}. But in our experiment, the situation is quite different for the nonpolar property of nitrogen and the weak terahertz wave intensity.
Here, we assume that the TWG from aligned nitrogen molecules is mainly determined by the ionization rates at the generating moment. According to these assumptions, the observed signals of HHG as well as TWG can be separated into two independent parts, namely the rotation of molecules and the dynamics of electrons. In this section, we present the deconvolution details on how to obtain the molecular structures.

\subsection{A. Non-adiabatical molecular alignment}

The linear symmetric top molecules can be aligned via `kicking' by the femtosecond laser pulse~\cite{Stapelfeldt2003}. Molecules with thermal distributions of rotational states, are coherently excited to form the broad rotational wave packet by the Raman transitions between different angular quantum numbers. This rotational wave packet revives the transient alignment condition several picoseconds after the alignment pulse.

The initial rotational wave packet of the molecules is given by the Boltzmann law at different angular momentum quantum number $J$ and its projection $M$.
Stimulating by the alignment laser field, the transition between different rotational eigenstates can be solved by the time dependent Schr\"{o}dinger equation, which is the Raman process coupling $J$ with $J,J\pm2$ states.

After the alignment pulse, molecules continuously rotate at different rotational eigenstates (neglecting the dispersion).
By coherently summing all occupied eigenstates, we are able to acquire the rotational wave packet at any moment after the alignment pulse in principle.
Due to the coherence between $J$ and $J\pm2$, the temporal evolution of HHG presents 1/4th revivals and can be characterized by the expectation value $\langle\cos^2\theta\rangle$ for nitrogen molecule~\cite{Itatani2004,Itatani2005PRL}, where $\theta$ is the angle between the molecular axis and the polarization axis of the alignment pulse. Recently, the discoveries on higher order fractional revival structures have been observed, corresponding to the coherence between $J$ and $J\pm4$, $\pm6$ even $\pm8$~\cite{Lock2012}.

In FIG.~3, we show the calculated time-dependent expectation values of $\cos^2\theta$, $\cos^4\theta$ and $\cos^6\theta$ for nitrogen molecules. The parameters for the calculation are obtained from Refs.~\cite{HerzbergNIST,Miller1990a,Miller1990b}, the rotational temperature is set at 90~K and the alignment laser is 60~fs, $0.7\times10^{14}$~W/cm$^2$. The general shapes of the expectation values of $\cos^{2n}\theta$ are similar, while the featured differences are the height of base lines and the appearance of high-order fractional revivals.

It is worth noting that the population on different rotational states is determined by the intensity of the alignment pulse and the initial rotational temperature~\cite{Mikosch2013, Lock2012}. Actually, the laser intensity is a distribution at the transversal surface along the propagation direction at the interaction volumes and the rotational temperature can be roughly estimated by the supersonic expansion property~\cite{Scoles1988atomic}.
The rotational temperature of molecules can be derived from the Fourier spectrum of the temporal evolutions of high harmonics in experiment~\cite{Yoshii09}. Meanwhile, by fitting the experimental results, a mathematical procedure has been proposed to precisely determine the two parameters~\cite{Mikosch2013}.

\subsection{B. Theory of HHG and TWG from aligned molecules}

\begin{figure}
\includegraphics*[width=3.4in]{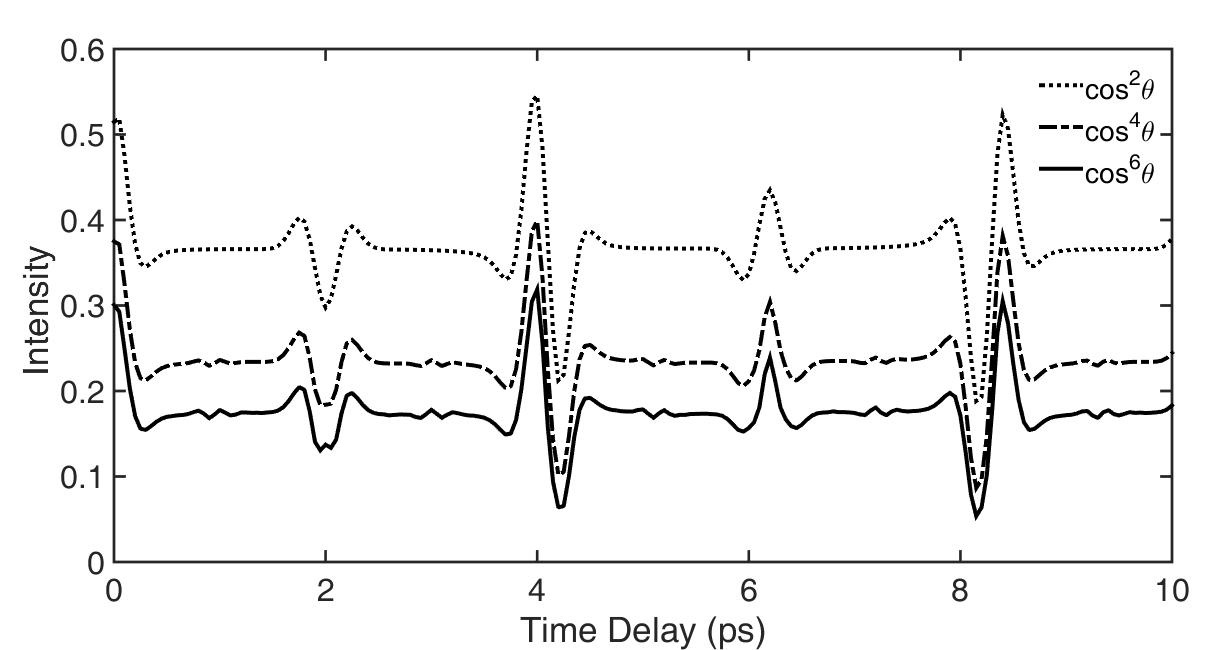}
\caption{\label{}From top to the bottom, the calculated $\langle\cos^2\theta\rangle$ (dotted lines), $\langle\cos^4\theta\rangle$ (dashed-dotted lines) and $\langle\cos^6\theta\rangle$ (solid lines) , as functions of the time delay between the alignment pulse and generation pulse.}
\end{figure}

In the following, we present the expressions of the alignment dependent HHG and TWG from molecules.

HHG has been proposed as a fascinating tool for the real-time image of the molecular orbital, due to the wavelength of electron is much shorter than the photon with the same energy~\cite{Haessler2011}. According to the QRS theory, angular yields of HHG from aligned molecules can be treated as a product of the rotational wave packet and the photorecombination cross section, which can be expressed by the Fourier component~\cite{LeAT2009},
\begin{equation}
I_{\rm HHG}\left( {\Omega,\theta} \right) \propto {\Omega ^4}{D^2}\left( {\Omega,\theta} \right),
\end{equation}
where $\Omega$ is the harmonic frequency, $D(\Omega, \theta)$ is the recombining dipole. Regardless the coherence between the microscopic wave packet $W(\Omega)$ and the recombination dipole moment $d(\Omega,\theta)$, the rescattering dipole $D_{\rm {QRS}}$ can be expressed as~\cite{JinCheng2012}
\begin{eqnarray}
D_{\rm {QRS}}\left( {\Omega ,\theta } \right) \approx \left(W\left( \Omega  \right)N{\left( \theta  \right)^{{1 \mathord{\left/{\vphantom {1 2}} \right.\kern-\nulldelimiterspace} 2}}}\right)d\left( {\Omega ,\theta } \right),
\end{eqnarray}
where $N(\theta)$ is the angular ionization rates.

The recombination process, $i.e.$ the third step for HHG, can be viewed as an inverse of photoionization process, thus the square of the recombination dipole moment $d^2(\Omega,\theta)$ is proportional to the angular differential PICS at energy $\omega$ ($\sigma_{\textit{\rm PICS}}$). Deriving from Eq.~(1) and (2), $I_{\rm HHG}$ can be rewritten as
\begin{equation}
I_{\rm HHG}\left( {\Omega ,\theta } \right) \propto {\Omega ^4}{W^2}\left( \Omega  \right){\sigma_{\textit{\rm PICS}}^{\vphantom{PICS}}}\left( {\Omega ,\theta } \right)N\left( {\theta}\right).
\end{equation}
It is worth noting that Eq.~(3) is for the case of single molecule, while the acquired HHG signals include the macroscopic effects. In experiments, the phase-matching for the short trajectory yields is realized by placing the orifice of gas jet after the focal to minimize the macroscopic effects.

Then we recall the theory of TWG in linear polarized dual-color laser field.
Considering the electron ionized at $t_i$, the angle-dependent drift current $j$ at any following moment $t$ can be expressed as
\begin{equation}
j(\theta; t_i,t)=-n_0(\theta; t_i)w(\theta; t_i)\int_{ti}^{t}\frac{eE(t')}{m_e}dt'.
\end{equation}
where $n_0(\theta; t_i)$ is the transient neutral molecule density, $w(\theta; t_i)$ is the transient ionization rates, $E(t')$ corresponds to the transient laser field, $e$ and $m_e$ is the constant of the electronic charge and mass. Here, we assumed the initial velocity of the tunneling electron is zero~\cite{ChenWB2015}. Therefore, the residual current $J_R$, associating to the electronic motion in the oscillating laser field, can be written as,
\begin{equation}
J_R(\theta, t)=\int_{-\infty}^{t}j(\theta; t_i,t)dt_i.
\end{equation}

Taking the form of the current model~\cite{KimKY2008,KimKY2009Review,YouYS2013PRA}, the far-field terahertz amplitude is proportional to the cycle-averaged electric current density $J_R(\theta, t)$,
\begin{equation}
E_{\rm THz}^{}(\theta)\propto\frac{d\langle J_R(\theta,t)\rangle}{dt}=\langle e v_d(t)n(\theta,t)\rangle,
\end{equation}
where $v_d(\theta)$ represents the drift velocity of an electron born at $t$, and $n(\theta,t)$ is the angular ionization rates at $t$. The cycle average of $n(\theta,t)$ is proportional to the angular ionization rates $N(\theta)$, therefore, $I_{\rm THz}^{}(\theta)$ is proportional to $N(\theta)^2$.

\subsection{C. Depict of the PICSs of molecules}
Replacing the angular ionization rates by the alignment dependent TWG, Eq.~(3) can be rewritten as
\begin{equation}
I_{\rm HHG}\left( {\Omega ,\theta } \right) \propto {\Omega ^4}{W^2}\left( \Omega  \right){\sigma_{\rm PICS}^{\vphantom{PICS}}}\left( {\Omega ,\theta } \right)\sqrt{{I_{\rm THz}}\left( \theta  \right)}.
\end{equation}
The first two items in the right side of Eq.~(9) only associate to the harmonic order. Therefore, the amplitude of angular PICS can be simplified by
expressing as,
\begin{equation}
\sigma _{\rm PICS}^{}(\Omega,\theta)\propto\frac{I_{\rm HHG}^{}(\Omega ,\theta)}{\sqrt{I_{\rm THz}(\theta)}}.
\end{equation}
Since $I_{\rm HHG}$ and $I_{\rm Thz}$ can be synchronizely measured, it is straightforward to obtain the differential PICS in one run, calibrating the AIR with alignment-angle dependent TWG.

It is worth noting that the minimums of the alignment dependent high harmonics reflect the harmonic phase jump~\cite{Lein2002PRA}, which is also the reason for the minimums of PICSs~\cite{LeAT2009}.
For two-center molecules, the total wave function of the molecules can be separated into two parts, namely the $\phi_{\rm sym}$ and $\phi_{\rm asym}$ corresponding to the symmetric and antisymmetric partial wave packet.
Therefore, it can be written as
\begin{eqnarray}
\Psi_{\rm MO}(\theta)&=&a_1\phi_{sym}(r)+a_2\phi_{\rm asym}(r),
\end{eqnarray}
with
$\phi_{sym}$ equaling to $\phi_1(r-{\rm R}/{2})+\phi_1(r+{\rm R}/{2})$ and $\phi_{\rm asym}$ equaling to $\phi_2(r-{\rm R}/{2})-\phi_2(r+{\rm R}/{2})$, both are determined by $\Psi_{\rm MO}(\theta)$.
Here, $a_1~(a_2)$ is the normalized population of the symmetric (antisymmetric) partial wave packet and R is the internuclear distance. The recombination dipole matrix element in velocity gauge is
\begin{eqnarray}
M_{\hat{d}}&=&\big{\langle}\Psi_{\rm MO}(\theta)|\hat{d}|e^{i\vec{k\mathstrut}\cdot \vec{r\mathstrut}}\big{\rangle}\nonumber \\
&=&2a_1k\cos(\frac{\vec{k}\cdot \vec{R}}{2})\big{\langle}\phi_1(r)|e^{i\vec{k\mathstrut}\cdot \vec{r\mathstrut}}\big{\rangle}\nonumber \\
&&-2ia_2k\sin(\frac{\vec{k}\cdot \vec{R}}{2})\big{\langle}\phi_2(r)|e^{i\vec{k\mathstrut}\cdot \vec{r\mathstrut}}\big{\rangle},
\end{eqnarray}
with k the wave number of the returning electron wave.
Denoting $q$ as the relative value between the plane-wave projections on the symmetric part $\phi_1$ and antisymmetric parts $\phi_2$, the interference term $T_{k,\theta}$ can be simplified as
\begin{equation}
T_{k,\theta}\propto\cos((kR\cos\theta)/2+\varphi),
\end{equation}
with $\cos\varphi=iqa_1/\sqrt{(iq)^2a_1^2+a_2^2}$.

The two-center interference (TCI) minimums (or maximums) appear when the interference term $T_{k,\theta}$ changes its sign.
Therefore, the corresponding alignment angle $\theta$ for two-center minimum (or maximum) satisfies the following equation,
\begin{equation}
kR\cos\theta=2n\pi+\pi-2\varphi,
\end{equation}
or
\begin{equation}
R\cos\theta=\frac{1}{\lambda_e}(n+\frac{1}{2}-\frac{\varphi}{\pi}),
\end{equation}
with $n$ the arbitrary integer. $\lambda_e$ is the wavelength of the recombining electron, whose kinetic energy is $E_{\Omega}-I_p$. When gradually varying the value of the returning electronic momentum $k$, the corresponding minimum angle $\theta$ will also shift, which has been theoretically and experimentally demonstrated for nitrogen~\cite{Zimmermann2005PRA,HuangYD2015}.

\subsection{D. Deconvolution of the HHG and TWG}
In this section, the primed and unprimed notations of the physical quantities indicate the laboratory frame and molecular frame, respectively.

The measured terahertz and high-harmonic yields in laboratory frame are the convolution of single molecule response and angular alignment distribution, and
we assume the signals are the incoherent sum of the single molecular yields from all solid angles~\cite{Yoshii2011}, $i.e.$,
\begin{eqnarray}
M\left({\alpha',t_{D}}\right){\propto} \int_0^{2\pi}d\varphi'\int_0^\pi{d\theta'S\left({\theta',\varphi';\alpha'} \right)}\nonumber\\
 \times\rho\left({\theta',\varphi';t_{D}}\right)\sin\theta'.
\end{eqnarray}
Here $\theta'$ ($\varphi'$) is the polar (azimuthal) angle about the polarization axis of the aligning pulse, $\alpha'$ represents the angle between the polarization directions of the alignment and generation pulses, $S\left({\theta',\varphi';\alpha'} \right)$ is the single molecular response and $\rho\left({\theta',\varphi';t_{D}}\right)$ is the angular distribution of the rational wave packet.

Typically, the deconvolution on single molecular response can be realized by tracing the time-delay dependent ($t_D$) or the alignment-angle dependent ($\alpha'$) yields~\cite{RenXM2013}.
For linear top symmetrical molecules, the single molecular yields $S$ only depends on the polar angle $\theta$ in the molecular frame, therefore we can simplify it as $S(\theta)$ and expand it by Legrand polynomials.
Meanwhile the coordinate transformation needs to be applied by using~\cite{Pavicic2007}
\begin{equation}
\cos\theta=\cos\alpha'\cos\theta'+\sin\alpha'\sin\theta'\sin\varphi'.
\end{equation}
The angular distribution of the rotational wave packet shares the cylindrical symmetry, therefore $\rho\left({\theta',\varphi';t_{D}}\right)$ can be simplified as $\varrho(\theta',t_{D})/{2\pi}$.
Based on the former deductions, the measured terahertz-wave and high-harmonic signal can be expressed as
\begin{eqnarray}
M\left({\alpha',t_{D}}\right){\propto} \int_0^{2\pi}d\varphi'\int_0^\pi{d\theta'S\left[{\theta(\theta',\varphi';\alpha')} \right]}\nonumber\\
 \times\varrho\left({\theta',t_{D}}\right)\sin\theta'.
\end{eqnarray}

Commonly, the single molecular response $S(\theta)$ can be estimated from the transition dipole or the TCI by fitting the experimental data~\cite{Lock2012}. While, $S(\theta)$ is a function of the alignment angle and can be expanded through. Therefore, if without the prior knowledge of the exact form of single molecular response, we can still reconstruct $S(\theta)$ by finding the suitable expansion parameters of the polynomials.

In specific, firstly we estimate the potential range of rotational temperature T and the alignment pulse intensity I and try different pairs of them to calculate the rotational wave packet $\varrho(\theta',t_{\rm al})$, with $t_{\rm al}$ the alignment moment. Meanwhile, we can expand the single molecular signal by the Legrand polynomials ${\rm P}_{i}(\cos\theta)$ with unknown expansion parameters $a_i^{}$,
\begin{equation}
S(\theta)=\sum_i a_i^{}\times {\rm P}_{i}(\cos\theta).
\end{equation}
In principle, the higher the upper limit of the expansion order is, the better fitting accuracy it will be.
Due to the symmetry of the nitrogen molecule demonstrated here, we only take the even orders of expansion into consideration.
Here, the expansion serious P$_{2j}(\cos \theta)$ with $j$ from 0 to 4 to depict nitrogen molecule are applied in reconstruction of $S(\theta)$. Given the complex numberical form of the Legrand polynomials, it is equivalent to use the Fourier expansion serious of $\cos^{2j}\theta$ in stead of P$_{2j}(\cos\theta)$.
Hence, we can rewrite Eq.~(16) as
\begin{equation}
S(\theta)=\sum_{j=0}^4 A_{2j}^{}\times(\cos^{2j}\theta),
\end{equation}
with $2j$ the corresponding expansion parameters.
The least square method is performed in searching for different pairs of the temperature and laser intensity $\{\rm I, T\}$ and the corresponding $A_{2j; \{\rm I, T\}}$, $i.e.$, comparing the experimental data at the alignment moment with the fitting results obtained from multiplying $S(\theta)$ with the $\varrho(\theta',t_{\rm al})$.

Secondly, all different pairs of $\{\rm I,T\}$ and the corresponding expansion parameters $A_{2j;\{\rm I,T\}}$ should be used to fit the angular emissions at the antialignment moment. The least squares method is applied to find the best fitting series from all the different pairs, denoting as $\{\rm I,T\}_b$ and $A_{2j;\{\rm I,T\}_b}$.

Thirdly, taking the obtained $A_{2j;\{\rm I,T\}_b}$ and ${\{\rm I,T\}_b}$ as the initial values to retrieve the temporal evolution of the different emissions. If we can reproduce the experimental results, all the fitting parameters are self consistent; if not, slightly adjustment of the fitting parameters should be try until convergent.

\section{IV. Experimental Results and Discussions}

\subsection{A. Temporal evolution of HHG and TWG}
\begin{figure}
\includegraphics*[width=3.4in]{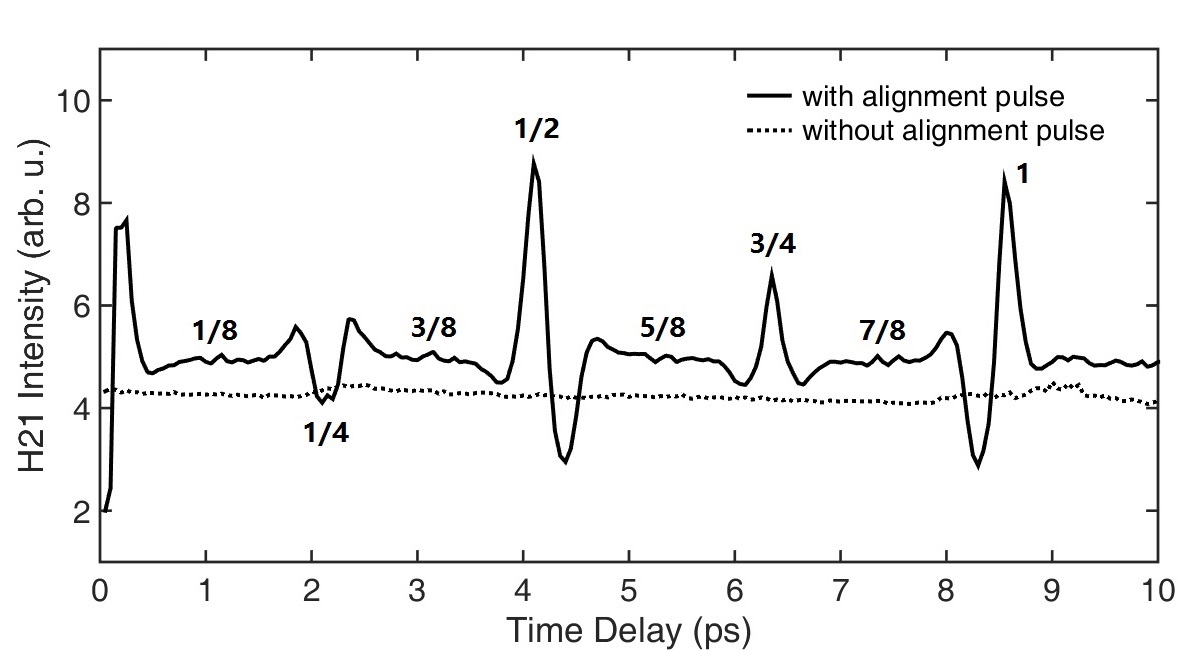}
\caption{\label{}The time-delay dependent modulations of the yields of 21st harmonic from nitrogen with (solid lines) or without (dashed lines) the alignment pulse. The fractional revivals are indicated at the corresponding time delays.}
\end{figure}

In FIG.~4, we show the modulations of the yields of 21st harmonic from nitrogen as a function of the time delay between the alignment and generation pulses. The integral multiplies of the $1/8 th$ revival for nitrogen are marked by their orders. We also show the harmonic signals without the alignment pulse in comparison, which are achieved by closing the beam shutter. Researches on nitrogen suggests that the temporal evolution of high harmonics generally follows the modulation of $\langle\cos^4\theta\rangle$~\cite{Ramakrishna2007}, which is in conformity with our results.

It is worth noting that the general structures of the expecting values for $\langle\cos^2\theta\rangle$, $\langle\cos^4\theta\rangle$ and $\langle\cos^6\theta\rangle$ are semblable, and it's enough to reproduce the majority structure of the high harmonics by the temporal evolution of $\langle\cos^4\theta\rangle$. However, in FIG.~4, we can still find some tiny oscillations around the odd times of $1/8 th$ revivals, which corresponds to the high-order coherence between different rotational states. These tiny structures may contribute to the high-order fractional revivals of nitrogen molecule, $i.e.$, the integral multiplies of the $1/12 th$ or the $1/16 th$ revivals, which have been reported on carbon dioxide~\cite{Lock2012}. However, to nitrogen molecule with 8.4 ps revival period (one-fifth of carbon dioxide's), high-order fractional revivals are too close to be separately distinguished.

The comparisons for TWG and HHG from nitrogen around the half revival moments are presented in FIG.~2 of Ref.~\cite{HuangYD2015}. The alignment and generation pulses are parallel and the temporal evolutions of HHG and TWG are similar to each other. Correlated measurements on ion fragments and high harmonics have been conducted for nitrogen~\cite{Kanai2005}, showing the consistent trend with ours. These results convincingly demonstrate our treatment on calibrating the angular ionization rates with alignment-dependent TWG is reliable.

\subsection{B. Shifting minima of the angular PICS}

\begin{figure}
\includegraphics*[width=3.5in]{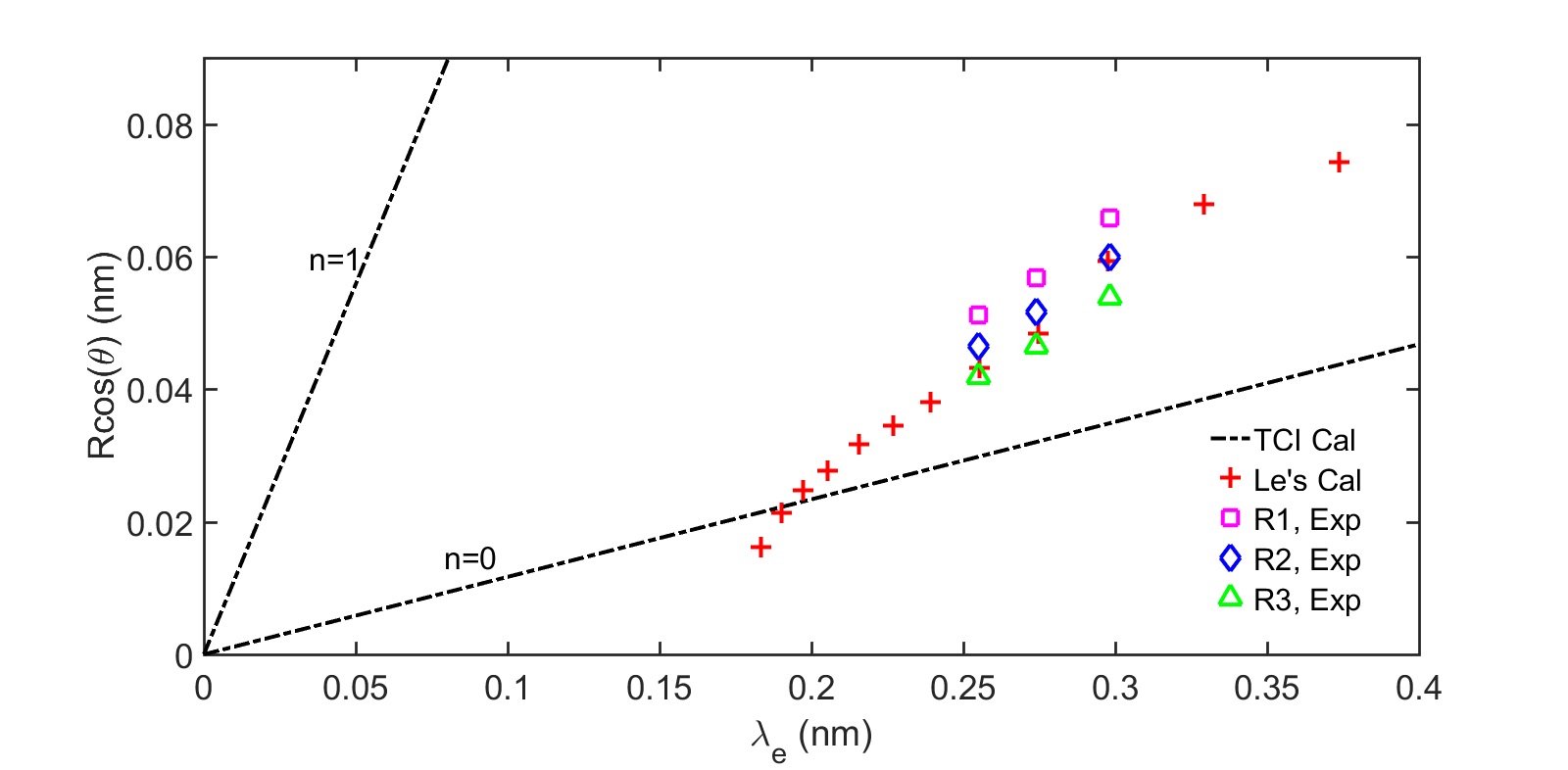}
\caption{\label{}The projection of the internuclear distance is plotted against the electron wavelength for nitrogen. The experimental deduced results are shown in violet squares ($R_1=0.121nm$), blue diamonds ($R_2=0.110nm$) and green triangles ($R_3=0.099nm$). The theoretical results based on the TCI model (use Eq.~(13)) are shown with black dashed-dotted lines. Le's calculations \cite{LeAT2009} on the minima of the PICS for nitrogen are presented (red crosses) in comparision.}
\end{figure}

According to the descriptions in Sec. III C, the angular differential PICS can be retrieved from the alignment-angle dependent TWG and HHG in molecular frame. While the measured signals are in lab frame, thus deconvolution is performed to obtain the alignment-angle dependent terahertz and harmonic yields of single molecule. The detailed methods are present in Sec. III D and the results of single molecular responses have been reported (see FIG.~3 in Ref.~\cite{HuangYD2015}).

For HHG from molecules, it's significant that intensity minimums can be observed spectrally or angularly\cite{Boutu2008,Kanai2005,Haessler2010}, which reflects the electron phase jump \cite{Lein2002PRA}. Therefore, figuring out the minimums and comparing between the experimental results and the theoretical predictions is indeed a must for understanding the molecular properties.

In FIG.~5, the experimental results are compared to the theoretical predictions of the minima positions. The results based on the TCI model, using Eq.~(13), exhibits the similar trend of our experimental results, $i.e.$, from harmonic order 21st to 25th, the alignment angle of minimum is gradually shifting to 90 degrees. According to the Hartree-Fock calculations by STO-3G basis, the HOMO of nitrogen can be linearly combined from the $2s$ and $2p_z$ atomic orbital (for convenience, we ignore the minor participation of $1s$ orbital), with the combination parameters 0.4 and 0.6, respectively.
The $2s$ and $2p_z$ orbital are symmetrical and the antisymmetrical, respectively, and the relative value q of the two partial wave packets equals to $1/\sqrt{3}i$. Thus, the additional phase shift $\varphi \sim 0.21\pi$ can be obtained. The black dashed-dotted lines in Fig.~5 show the situations when n equals 0 and 1.

The calculation results on the angular PICS of nitrogen are also presented \cite{LeAT2009}, exhibiting complicated relationships between the projection of the internuclear distance and the electron wavelength \cite{Zimmermann2005PRA}. The wavelength in Ref.~\cite{LeAT2009} is the `effective' electron wavelength, which acquired from $E_{k}^{\rm eff}=E_{\Omega}$. Here, we shift it by removing the ionization potential of nitrogen, $i.e.$, $E_k=E_\Omega-I_p$.

The projections of the internuclear distance associating to the our experimental results are presented to compare with the calculation results, with bond length of nitrogen $R_1=0.121nm$, $R_2=0.110nm$ and $R_3=0.099nm$ \cite{Blaga2012}. $R_2$ is the equilibrium N-N distance for neutral nitrogen, which fits better for Le's calculations. Whereas, the experimental results show discrepancy with the simple TCI calculations in the region we concerned and the angles for the simple TCI model extend about $10$ degrees than our deduced results. The differences between TCI results and the experimental results can be mainly attributed to the uncertainty of the combination parameters of the $2s$ and $2p_z$ atomic orbital and the variation of the additional phase $\varphi$ \cite{Zimmermann2005PRA}.

\subsection{C. Optimal phase for TWG}
\begin{figure}
\includegraphics*[width=3.4in]{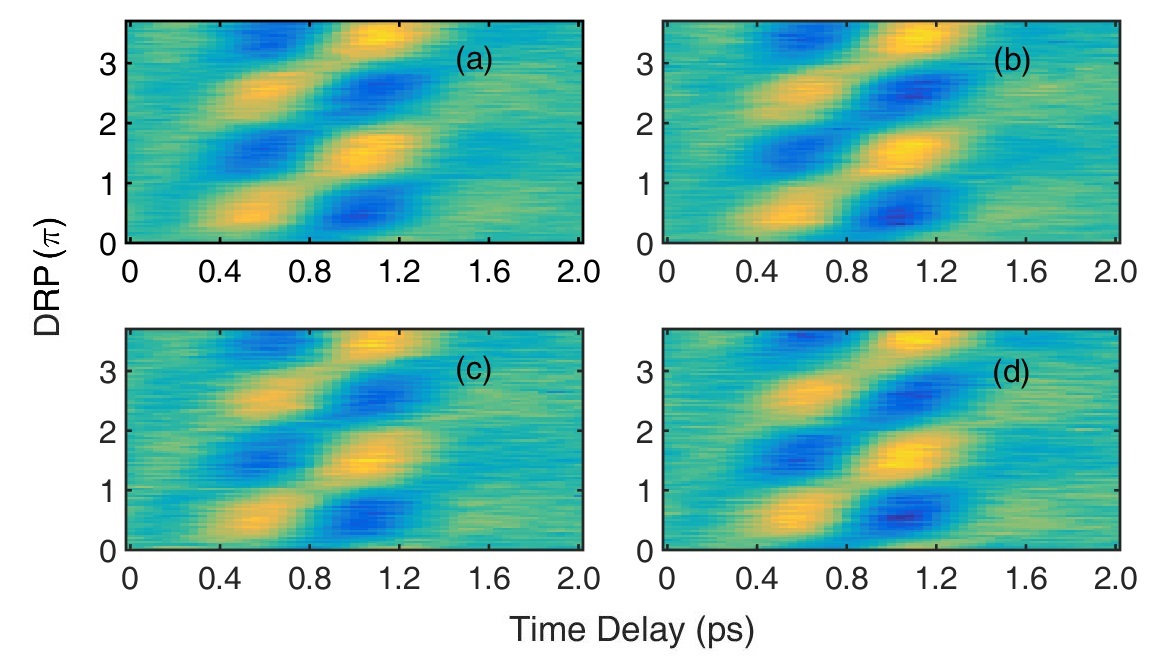}
\caption{\label{}THz time-dependent amplitude versus the DRP from nitrogen. (a) and (c) are the experimental results from aligned and antialigned nitrogen molecules, respectively; (b) and (d) are the corresponding the isotropic results without alignment pulses.}
\end{figure}

By varying the transmission length of the wedges, we obtain the DRP-modulated TWG from aligned nitrogen, while no observable variation on the optimal phase is found at the accuracy of about $50$ as in our experiment. In FIG.~6, we show the THz waveform versus the DRP. The inverse of the THz waveform for different DRP indicates the inversion of the current direction. By continuously turning the shutter on and off, the terahertz waveform is traced from nitrogen molecules with or without the alignment pulses. Recently, as is reported by Chen~\cite{ChenWB2015}, theoretical work on aligned molecules have present the difference of the optimal phase for TWG is about 17 as between the 0 degree aligned and 90 degree aligned model $H_2^+$. To observe this tiny difference is definitely a tough challenge to the current experimental methods.
In FIG.~7, we show the modulations of the THz intensity versus the DRP of argon and isotropic nitrogen taken under the same experimental conditions. The nearly overlapped lines on modulations indicate the similar manner of electrons from nitrogen and argon, which is determined by the combination of Coulomb potential and the focused dual-color fields.

Argon atom shares the closing ionization potential to the nitrogen molecule, which makes it a suitable comparision in studying THz generating process.
If the electrons, no matter the escaping or the rescattering ones, are far from the nuclei which is screened by other electrons, the effect of the Coulomb potential works like a positive core, only slightly adjust the trajectory. Therefore, to small molecules such as nitrogen, the TWG is almost generated at the same time from different aligned angles.
Unless the gating of dynamics is enhanced to a much more precise resolution \cite{Eckle2008}, it is hard to figure out the minor difference on the emission time for TWG. That's the reason why no difference is found in our experiments on the optimal phase for the aligned, antialigned, isotropic nitrogen molecules and the argon atoms.

\begin{figure}
\includegraphics*[width=3.6in]{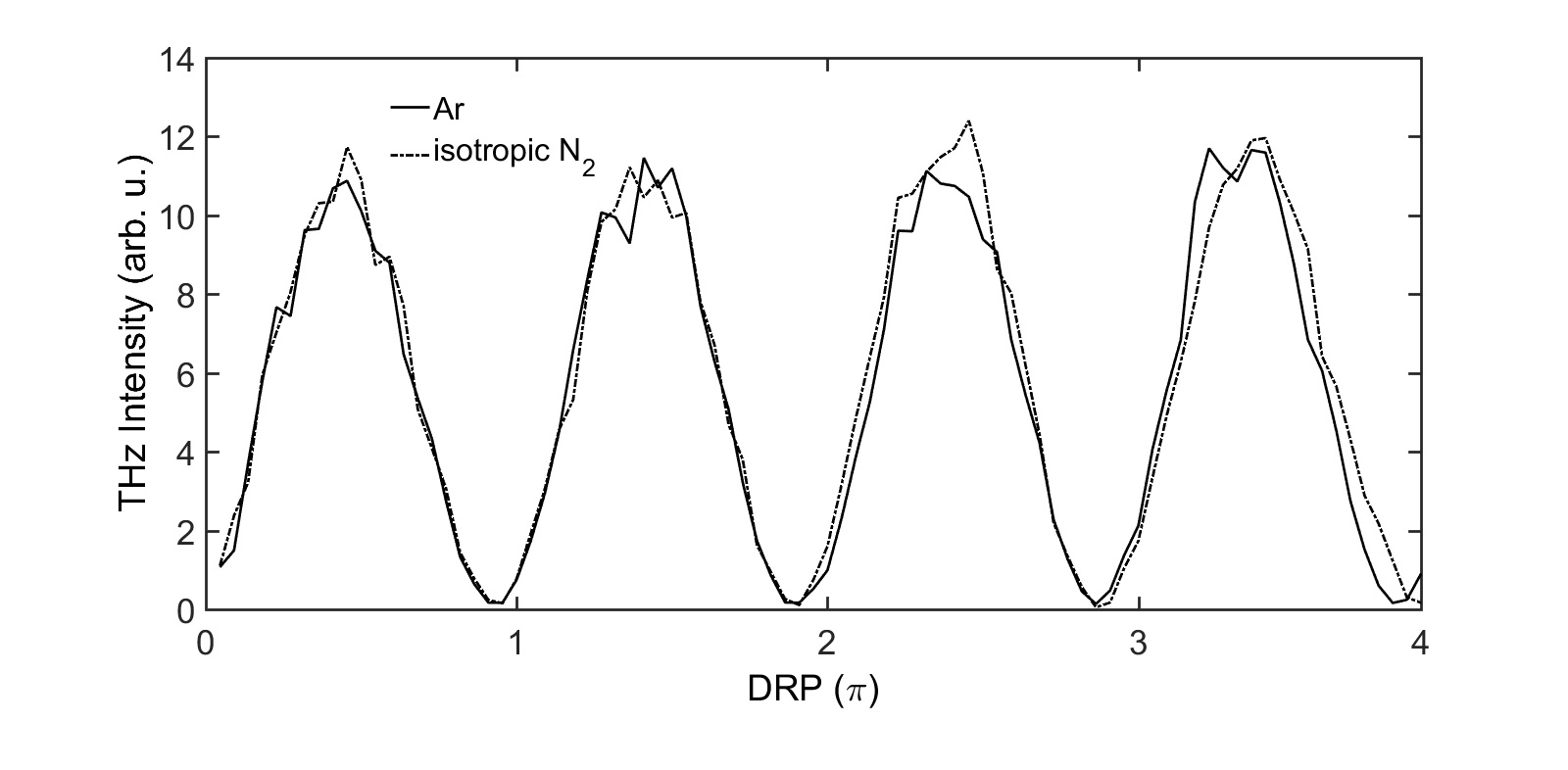}
\caption{\label{}THz intensity versus the relative phase between the $\omega$ and $2\omega$ pulses with parallel polarization from isotropic nitrogen molecule (black solid lines) and argon atom (grey dashed lines).}
\end{figure}

\section{V. Conclusions and Outlook}

In summary, we present the experimental and theoretical details for the HATS from aligned N$_2$ molecules.

On one hand, HATS, saving separate measurements on AIR, is introduced as a brand-new method in encoding the structure and dynamics of molecule in strong laser fields.
Due to the importance of AIR for molecules, applying alignment-dependent TWG to calibrate the ionization process is indeed a supplement to the contemporary approaches both in theory and experiment.
Meanwhile, the qualitative agreements on the minima angles of the PICS between the theoretical and experimental results show the potential applications in investigating the structure and emission properties of molecules, which can be applied on much more complicated molecules.

On the other hand, we compare the DRP modulated TWG from molecules and atoms. Within our experimental precision ($50$ as), no significant difference on the optimal phase for TWG is observed between nitrogen molecules and argon atoms, no matter aligned or not. This means the Coulomb potential of nitrogen only sightly adjusts the trajectory of the ionized electrons.
To exhibit the influence of Coulomb focusing on electron dynamics during TWG process, attempts on larger molecules and enhancement on the detection precision are the predictable ways, which challenge the experimental methods.

\section{Acknowledgment}
This work is supported by the National Basic Research Program of China (973 Program) under Grant No. 2013CB922203, and the NSF of China (Grants No. 11374366 and 11474359). Y. H. thanks to the constructive discussions with Wenbo Chen, Bin Zhang, Wenpu Dong, Jinlei Liu, Lu Liu and Quan Guo.

%

\end{document}